\documentclass[runningheads]{llncs}
\usepackage{graphicx}
\usepackage{bm}
\usepackage{float} 
\newcommand{\squeezeup}{\vspace{-3mm}}
\newcommand{\squeezeupCap}{\vspace{-7mm}}
\begin{document}
\title{Recurrent Neural Network Learning of Performance and Intrinsic Population Dynamics from Sparse Neural Data}
\titlerunning{RNN Learning of Performance and Intrinsic Population Dynamics}
%
\author{Alessandro Salatiello\inst{1,2} \and
Martin A. Giese\inst{1,2}}
\authorrunning{Salatiello, A. et al.}
%
\institute{
Section for Computational Sensomotorics, Department of Cognitive Neurology, 
Centre for Integrative Neuroscience \& Hertie Institute for Clinical Brain
Research, University Clinic Tübingen, Germany \\
\and
International Max Planck Research School for Intelligent Systems, Tübingen, Germany\\
\email{\{alessandro.salatiello, martin.giese\}@uni-tuebingen.de}}
\maketitle              
\begin{abstract}
Recurrent Neural Networks (RNNs) are popular models of brain function. The typical training strategy is to adjust their input-output behavior so that it matches that of the biological circuit of interest. Even though this strategy ensures that the biological and artificial networks perform the same computational task, it does not guarantee that their internal activity dynamics match. This suggests that the trained RNNs might end up performing the task employing a different internal computational mechanism, which would make them a suboptimal model of the biological circuit. In this work, we introduce a novel training strategy that allows learning not only the input-output behavior of an RNN but also its internal network dynamics, based on sparse neural recordings. We test the proposed method by training an RNN to simultaneously reproduce internal dynamics and output signals of a physiologically-inspired neural model. Specifically, this model generates the multiphasic muscle-like activity patterns typically observed during the execution of reaching movements, based on the oscillatory activation patterns concurrently observed in the motor cortex. Remarkably, we show that the reproduction of the internal dynamics is successful even when the training algorithm relies on the activities of a small subset of neurons sampled from the biological network. Furthermore, we show that training the RNNs with this method significantly improves their generalization performance. Overall, our results suggest that the proposed method is suitable for building powerful functional RNN models, which automatically capture important computational properties of the biological circuit of interest from sparse neural recordings.


\end{abstract}
\section{Introduction}
Recurrent Neural Networks (RNNs) are a common theoretical framework for the modeling of biological neural circuits. The rationale behind this choice is that RNNs and biological neural systems share many essential properties. First, computations are performed through complex interactions between local units; second, these units act in a parallel and distributed fashion, and third, they communicate through recurrent connections \cite{sussillo2014neural}.
\\ \indent
Moreover, recent work \cite{mazor2005transient,churchland2012neural,mante2013context,elsayed2016reorganization,williams2018unsupervised,pandarinath2018inferring,chaudhuri2019intrinsic,gallego2020long} has re-emphasized the dynamical systems perspective of brain computation \cite{schoner1988dynamic}. This has renewed interest in using RNNs, long known to be universal approximators of dynamical systems \cite{doya1993universality}. In brief, this perspective recognizes the intrinsic low-dimensionality of the activity patterns displayed by populations of neurons, and has been focusing on identifying and studying the low-dimensional projections of such patters, along with the {\it dynamic rules} that dictate their evolution in time; see \cite{shenoy2013cortical,cunningham2014dimensionality,yuste2015neuron,saxena2019towards} for a comprehensive review of such principles.
\\ \indent
RNNs have been used extensively in neuroscience \cite{Daya_2001_book} and have provided fundamental insights into the computational principles 
of different brain functions, including working memory \cite{machens2005flexible}, decision making \cite{mante2013context}, behavior timing \cite{wang2018flexible}, and motor control \cite{hennequin2014optimal}. Nevertheless, the design of such networks and the determination of their parameters in a way that they capture the input-output behavior and the underlying internal computational mechanism of the biological circuit of interest remains an unsolved problem. 
\\ \indent
When there is a general understanding of the dynamics necessary to implement a specific function, RNNs can be designed {\it by hand} to generate the required dynamics (e.g. \cite{matsuoka1987mechanisms,machens2005flexible,pollock2019engineering}). However, in most cases, this understanding is lacking, and the network design requires data-driven parameter training. The most popular training strategies fall into two categories: {\it task-based} (also referred to as {\it goal-driven}) and {\it activity-based} training (see \cite{saxena2019towards} for a review).
\\ \indent
The first strategy is to train the RNN to perform the (supposed) computational function of the target biological circuit (e.g. see \cite{mante2013context,yamins2014performance}). This approach ensures that the biological and model networks perform the same computational task. However, it does not guarantee that the they employ the same internal computational mechanism to perform the task \cite{williamson2019bridging,maheswaranathan2019universality}. As a matter of fact, the population dynamics of the trained RNN frequently differ from those of the target biological network. The second strategy is to train the RNN to directly reproduce the recorded neuronal activity patterns (e.g. see \cite{iclr:BattyMBHSLCP17,kim2018learning}). The resulting models typically explain high fractions of the neural recordings' variance. However, in many cases, it remains unclear to which degree such networks perform the global underlying computational task. 
\\ \indent
In this paper, we introduce a novel training strategy that combines the strengths of the previously discussed approaches. Specifically, our training strategy makes it possible to train an RNN to perform the same computational task that a target biological network performs, while at the same time ensuring that the internal network dynamics are highly similar to those of the modeled neural system.
This was accomplished by exploiting {\it full-FORCE learning} \cite{depasquale2018full}.
This algorithm has been shown to be effective for training RNNs to internally generate given target dynamics. In \cite{depasquale2018full}, the authors showed that when appropriately designed, such target dynamics can significantly improve RNN performance. However, the target dynamics used in this work were specified by rather simple {\it hint signals} (ramp and step functions), which were carefully designed to solve the task at hand and had no direct relationship with the activity patterns of real neurons.
\\ \indent
Here, we extend this work and demonstrate that this training algorithm can be used to embed biologically-relevant dynamics in RNNs that perform biological functions, using sparse neural recordings. For this purpose, we first generate ground-truth data using a physiologically-inspired neural model. Then, we train an RNN to simultaneously reproduce both the input-output behavior and the internal dynamics of this model.
Critically, we show that the RNNs trained with our method achieve superior generalization performance and are thus able to capture important computational properties of the target neural model. 

\section{Methods}
\subsection{Instructed-Delay Center-Out Reaching Task}
\begin{figure}[t]
\includegraphics[width=\textwidth]{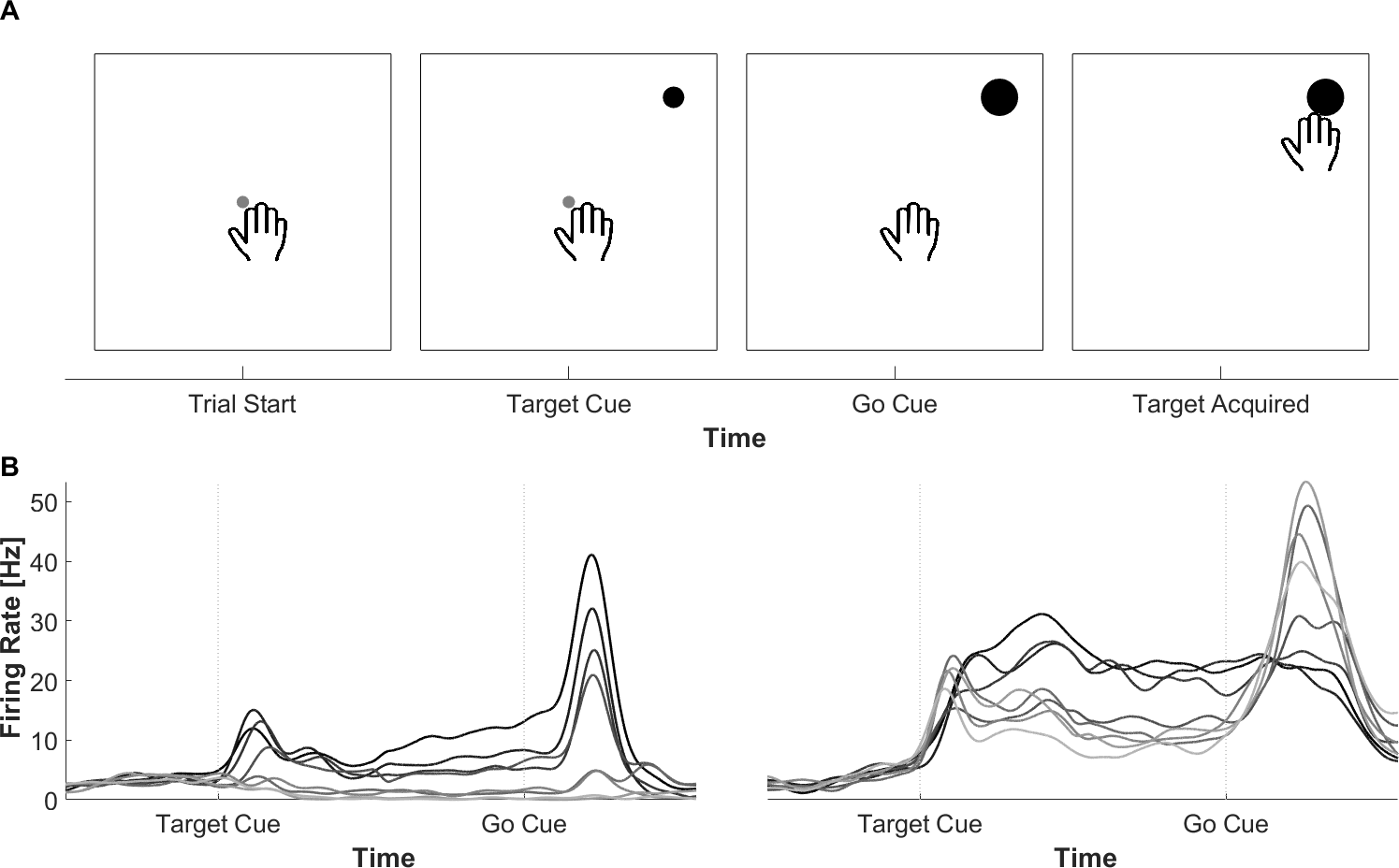}
\squeezeupCap
\caption{\textbf{Simulated task and typical motor cortical activation patterns.\quad \quad \quad}
 \textbf{(A)} Center-out reaching task. Grey circle: starting position; black circle: target. \textbf{(B)} Firing rates of two representative neurons in dorsal premotor cortex (PMd) during a 7-target reaching task \cite{churchland2010stimulus}; each trace represents the across-trial mean for one of the seven conditions. The data, retrieved from the online repository \cite{churchRep}, were separately averaged locked to Target Cue and Go Cue.}\label{fig1}
\end{figure}

To test the effectiveness of the proposed method, we applied it to learn a neural control model for the center-out reaching task \cite{georgopoulos1981spatial}, one of the most studied tasks in motor control.
This task (Fig. \ref{fig1}A) requires reaching movements from a central starting position towards one of $n_{tar}$  targets located at a fixed distance from it. First, participants are presented with a {\it Target Cue}, which indicates which target to reach to. Subsequently, after a variable delay period, they are instructed to perform the reach by an additional {\it Go Cue}. Variation of this delay period allows the study of the cortical dynamics associated with movement planning.
\\ \indent
Studies in non-human primates have shown that concurrently recorded neurons in the motor cortex exhibit reproducible activity patterns during the different phases of the reach (see \cite{shenoy2013cortical} for a review). In brief, during the delay period, the firing rates slowly transition from a baseline resting value to a steady-state value that varies with reach direction (Fig. \ref{fig1}B). Subsequently, after the Go Cue, the firing rates transition to an oscillatory regime that levels off before the target is reached (Fig. \ref{fig1}B). The dynamical systems perspective of the motor cortex \cite{shenoy2013cortical} postulates that the activity observed during the delay period sets the initial state for subsequent autonomous oscillatory dynamics. Such dynamics generate
control signals that are transmitted through lower motor centers to the muscles.
\\ \indent
Furthermore, studies in both humans and non-human primates have recorded electromyographic signals (EMGs) from relevant upper-arm, shoulder and chest muscles, and have consistently observed multiphasic activity patterns during the movement phase of the reaching task. Such patterns can be explained by the superposition of few time-varying {\it muscle synergies} \cite{d2003combinations,flash2005motor}, which capture the coordinated activation of muscle groups.
\subsection{Generation of Ground-Truth Data}
In order to validate the proposed method, we generated ground-truth data using a custom biologically-inspired generative model. Such a model is able to reproduce the activity patterns typically observed in motor cortex and muscles during reaching movements.
The model combines a Stability-Optimized Circuit (SOC) network \cite{hennequin2014optimal} for the generation of motor cortex-like activity, with a synergy-based \cite{flash2005motor} model for the generation of multiphasic EMG-like activity.

Specifically, we first generated EMG-like patterns for $n_{tar} = 90$ target angles; we then trained the SOC-network to generate the EMG patterns as stable solutions of its neural dynamics, starting from different initial conditions. Importantly, this type of network was shown to be able to autonomously generate oscillatory activation patterns that closely resemble the ones observed in the motor cortex during the movement phase of the reaching task.
\squeezeup
\subsubsection{Generation of EMG-like Patterns Through a Synergy-Based Model.}
\begin{figure}[t]
\includegraphics[width=\textwidth]{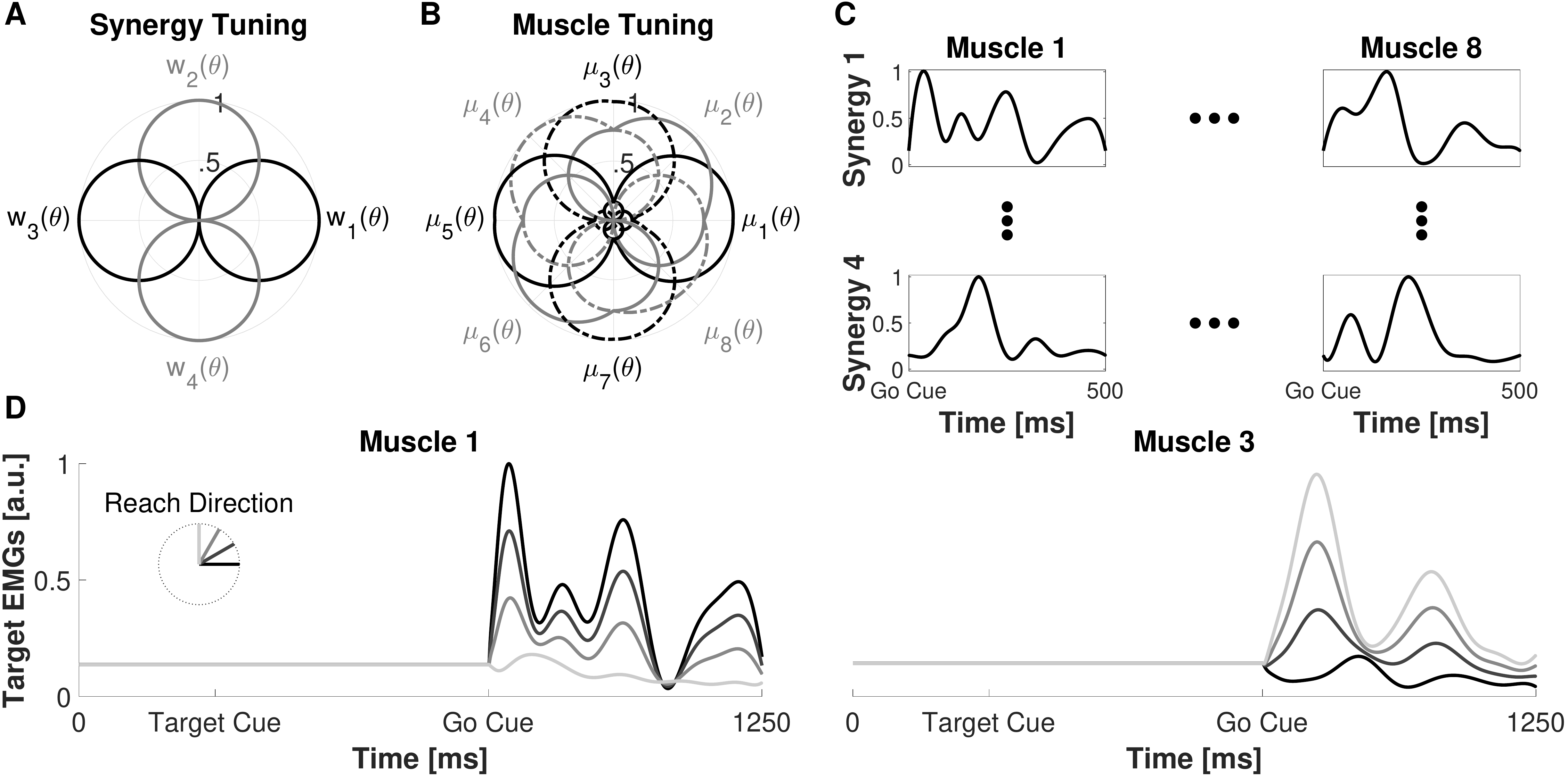}
\squeezeupCap
\caption{\textbf{Synergy-based generative model}. \textbf{(A)} Direction tuning of muscle synergies $w_s(\theta_{tar})$. \textbf{(B)} Direction tuning of individual muscles, computed maximizing over time the muscle activities (i.e. $ \boldsymbol{\mu}_i (\theta_{tar}) = \max\limits_{t} {\bf m}_i (t,\theta_{tar})$ ). \textbf{(C)} Synergies' activation patterns $\mathbf{s}_s(t)$. \textbf{(D)} Muscle activities (EMGs) of two representative muscles during reaches to four target directions. Traces are color coded based on the target direction. Note the smooth modulation of EMGs' amplitude and shape as the reaching direction changes.} \label{fig2}
\end{figure}

We generated time-varying EMG patterns only for the $t_{mov}=500$ samples of the movement phase of the reaching task. During the remaining period, each EMG signal was set to a muscle-specific resting level. The model assumes the existence of $n_{mus}=8$ muscles that mainly actuate the directions $ \theta_m = 0, \pi/4, ..., 7\pi/4$. We assume that these muscles are driven by $n_{syn}=4$ muscle synergies with preferred directions $\theta_s = 0, \pi/2, ..., 3\pi/2$.
\\ \indent
Each synergy is characterized by a time-dependent activity vector $\mathbf{s}_s(t)$ with $n_{mus}$ components; this vector specifies muscle-specific and time-varying activation patterns (Fig. \ref{fig2}C.). These patterns are generated as samples (normalized between $0$ and $1$) from a Gaussian process with kernel $K(t,t') = K_{se}(t,t') K_{ns}(t,t')$.
This kernel is given as product of a squared-exponential kernel $K_{se}(t,t') = e^{-(t-t')^2/{2 \zeta^2}}$ and a non-stationary kernel $K_{ns}(t,t') = E(t) E(t')$, where $E(t)  = (t/\sigma) e^{-(t/\sigma)^2/4}$. This specific kernel, originally proposed by \cite{stroud2018motor}, has been shown to generate samples that closely resemble the multiphasic EMG signals observed during reaching.
 \\ \indent
The muscle activities at time $t$ for target direction $\theta_{tar}$ (Fig. \ref{fig2}D) are given by a weighted sum
of the synergy activations according to the equation:
\begin{equation}
\mathbf{m}(t,\theta_{tar}) = \sum_{s=1}^{n_{syn}} w_s(\theta_{tar}) {\bf A}_{s}  \mathbf{s}_s(t)
\end{equation}
The function $w_s(\theta_{tar}) = [ \cos(\theta_{tar} - \theta_{s})]_+  $ defines the direction tuning
of the synergies, while the matrix ${\bf A}_s = {\rm diag}([\cos(.9(\theta_s - \theta_1))]_+,...,[\cos(.9(\theta_s - \theta_{n_{mus}}))]_+)$ defines the synergy tuning of the muscles, and thus determines how much each muscle $m$ is activated by the synergy $s$ (Fig. \ref{fig2}A,B.).

\squeezeup
\subsubsection{Generation of Motor-cortex-like Activity Patterns Through SOC Network.}
SOC networks \cite{hennequin2014optimal} are  non-linear, rate-based recurrent networks that respect Dale's principle.
The specific algorithm used to train these networks renders their internal network dynamics remarkably similar to the activity patterns observed in the motor cortex during reaching movements. Since these networks generate motor-cortical-like data, in the remainder of the paper, we will refer to the trained SOC network as the {\it biological network}. In this section, we explain how we trained an SOC network of 500 neurons in order to generate the reach-related EMG signals (Fig. \ref{fig2}D).
\\ \indent
The training comprises five steps (see \cite{hennequin2014optimal} for details). First, we initialized the recurrent connectivity matrix $\bf W$ randomly, ensuring that its spectral radius was $R=10$; this choice generates a strongly chaotic network behavior. Second, we made the network linearly stable by fine-tuning its inhibitory connections through an iterative algorithm, which minimizes the {\it smoothed spectral abscissa} of $\bf W$. Third, we found the four (orthogonal) initial conditions that generated the strongest transient responses in the network. These responses were arbitrarily associated with the {\it main target directions} $0, \pi/2, \pi, 3\pi/2$. Fourth, we linearly interpolated the four initial conditions on a quadrant-by-quadrant basis to generate initial conditions that are appropriate to reach to targets at intermediate angles. These directions define $n_{tar}$ initial conditions, which all result in strongly oscillatory dynamics. Fifth, we learned by linear regression, across all the conditions, a single readout matrix that maps the activity of the network onto the EMG activity patterns $\mathbf{m}(t,\theta_{tar})$. 
\subsection{Learning of the RNN Model using the Full-FORCE Algorithm}
The goal of our work is to train rate-based RNNs to generate the multiphasic EMG-like signals produced by the biological network, while at the same time generating similar internal activation dynamics. If the training procedure is successful, we end up {\it embedding} the activation dynamics of the biological network in the trained RNNs. Therefore, in the remainder of the text, we will refer to these trained RNNs as {\it embedder networks}. 
These networks obey the dynamics:
\begin{equation}
\tau \frac{d\mathbf{x}(t)}{dt} = -\mathbf{x}(t) + {\bf J}H(\mathbf{x}(t)) + {\bf L} {\bf u}(t)
\label{eq:embnn}
\end{equation}
The term $H(\mathbf{x}(t)) = \tanh(\mathbf{x}(t))$ represents the instantaneous firing rate as a function
of the membrane potential vector ${\bf x}(t)$. The output signals
that approximate the EMG-like signals $\mathbf{m}(t,\theta_{tar})$ are given by the vector 
$\mathbf{y}(t) = {\bf K} H(\mathbf{x}(t))$, where $\bf K$ is a read-out matrix. 

The input signal vector $\mathbf{u}(t)$ has 12 dimensions and selects the relevant reaching direction. 
Its components are given by smooth ramping signals that ramp up between the Target Cue and the Go Cue,
and then suddenly go back to zero to determine movement initiation. Their steady-state amplitudes, reached a few (40) ms before the Go Cue, code for reach direction. To determine such steady-state values, we used a procedure that mirrors the one we adopted to assign initial conditions to the biological network. Specifically, we first arbitrarily assigned 
12-dimensional random vectors to each of the {\it main angles} $0$, $\pi/2$, ..., $3\pi /2$, where their components were randomly drawn (with replacement) from the interval $[-0.6, 0.6]$. We then determined the amplitude components for 
intermediate angles by linearly interpolating the amplitudes associated to the main angles, on a quadrant-by-quadrant basis. 

As training data, we used simulated EMG and neural activity data for reach directions
0, 8, 16, ..., 352 deg. As test data we used the corresponding data for reach directions 4, 12, ..., 356 deg,
ensuring disjoint training and test sets. Importantly, the biological network generates data only for the 
movement phase of the reaching task, and we used only these data to embed the dynamics of the biological network in the embedder network. 
However, the embedder network had to generate the EMG signals and appropriate dynamics for the whole task, including the baseline and preparatory period. Critically, after training, also the embedder network produces the EMG signals through its autonomous dynamics, since the 
input signals ${\bf u (t)}$ return to zero right after the Go Cue to signal movement initiation. 
\squeezeup
\subsubsection{Full-FORCE Learning Algorithm.} We used the full-FORCE algorithm \cite{depasquale2018full} to learn the
recurrent weights ${\bf J}$ and the read-out weights $\bf K$. This algorithm constructs an additional {\it auxiliary network}
that receives a subset of the desired internal network activities as an additional input signal $\mathbf{u}_{hint}(t)$.
In our case, this hint signal is defined by the firing rates of the neurons of the biological network we assume to be able to record from.
The dynamics of the auxiliary network are given by:
\begin{equation} \label{eqHelper}
\tau\frac{d\mathbf{x^D}(t)}{dt} = -\mathbf{x^D}(t) + {\bf J}^D H(\mathbf{x^D}(t)) + {\bf L}\mathbf{u}(t) + {\bf B} \mathbf{u}_{hint}(t)
+ {\bf C} \mathbf{m}(t,\theta_{tar})
\end{equation}
where ${\bf J}^D$, ${\bf L}$, ${\bf B}$, and ${\bf C}$ are fixed random matrices that are not trained
The role of the auxiliary network is to generate dynamics that are driven by the input variables $\mathbf{u}(t)$, $\mathbf{u}_{hint}(t)$ and $\mathbf{m}(t,\theta_{tar})$\footnote{These input terms are strong enough to suppress chaotic activity in the network \cite{depasquale2018full}}. The core idea of the algorithm is to train the recurrent weights ${\bf J}$ of the embedder network (\ref{eq:embnn}) so that this generates signals related to such input terms, through its recurrent dynamics. This can be achieved by minimizing the error function
\begin{equation}
    Q({\bf J}) = \langle |{\bf J}H(\mathbf{x}(t)) - {\bf J}^DH(\mathbf{x^D}(t)) - {\bf B}\mathbf{u}_{hint}(t) - {\bf C} \mathbf{m}(t,\theta_{tar})  |^2 \rangle
\end{equation}
using a recursive least squares algorithm \cite{Haykin:2002}\footnote{Following \cite{depasquale2018full} we included an $L2$ regularization term for ${\bf J}$.} 
\squeezeup
\subsubsection{Embedding the Dynamics from Sparse Neuronal Samples.}
In real experiments, typically, one is able to record the activity of only a small fraction of neurons of the biological circuit of interest. It is thus important to assess whether one can efficiently {\it constrain} the dynamics of the embedder network using only the activities of small subsets of neurons sampled from the biological network. These 'recorded' neurons define the dimensionality of the additional input term $\mathbf{u}_{hint}(t)$\footnote{To ensure a sufficiently strong effect on the embedder network, the hint signals were scaled by factor 5. In addition, the final 110 samples of such signals were replaced by a smooth decay to zero, modeled by spline interpolation. This ensured that the activities go back to zero at the end of the movement phase.}.
Specifically, in our experiments, we constrain the embedder network dynamics using different percentages of neurons $\gamma = $ 0, 1, 2, 3, 10, and 100 \%, sampled from the biological network. Clearly, when $\gamma=0 \%$, we are effectively not embedding the dynamics.
Furthermore, we investigated whether an embedder network of a size $\eta$ smaller than the one of the biological network is still able to perform the task accurately and to learn its dynamics. For this purpose, we trained embedder networks of $\eta =$ 100, 200, and 500 neurons.

\subsection{Quantification of Embedder Network Performance}
\subsubsection{Features of Interest.} To quantify how well the embedder network is able to reproduce the EMG signals, we computed the generalization
error (GE), defined as the normalized reconstruction error on the test set, averaged across muscles and tasks.
To quantify the similarity between the population activity dynamics of the biological ($D_B$) and embedder networks ($D_{E}$), we computed their singular vector canonical correlations (CC){\footnote{We restricted our analysis to the first five singular vector canonical variables, which on average captured $>92\%$ of the original data variance.}}\cite{raghu2017svcca}. Moreover, as a baseline, we also computed the singular vector canonical correlations between the population dynamics of the biological network and those of an untrained RNN of 100 neurons with random connections ($D_0$).
\squeezeup
\subsubsection{Statistics.} To measure the overall effect of embedding the dynamics on GE and CC, we fitted the linear regression models  
$GE = \beta_0 + \beta_1 EoD + \beta_2 \eta$ and $CC = \beta_0 + \beta_1 EoD + \beta_2 \eta$, where $EoD$ is an indicator variable that takes on the value 1 whenever $\gamma>0$. Note that $\eta$ was included in the models since it is a potential confounder. Moreover, to quantify the effect of using different percentages of sampled neurons to embed the dynamics on GE and CC, we also fitted the linear regression models $GE = \beta_0 + \beta_1 \gamma + \beta_2 \eta$ and $CC = \beta_0 + \beta_1 \gamma + \beta_2 \eta$, considering only the networks with $\gamma>0$.
Finally, to summarize the results of the canonical correlation analysis, we also computed the {\it activation similarity index} (ASI), which is a normalized measure of the similarity between the activation dynamics of the biological network ($D_B$) and embedder network ($D_E$). This index is defined as
\begin{equation}
    ASI_{D_E} = \frac{mCC(D_E, D_{B}) - mCC(D_0, D_{B}) }{1 - mCC(D_0, D_{B})}
\end{equation}
where $mCC(D_i,D_{j})$ denotes the average CC between the activation dynamics $D_i$ and $D_j$.

\begin{figure}[t]
\includegraphics[width=\textwidth]{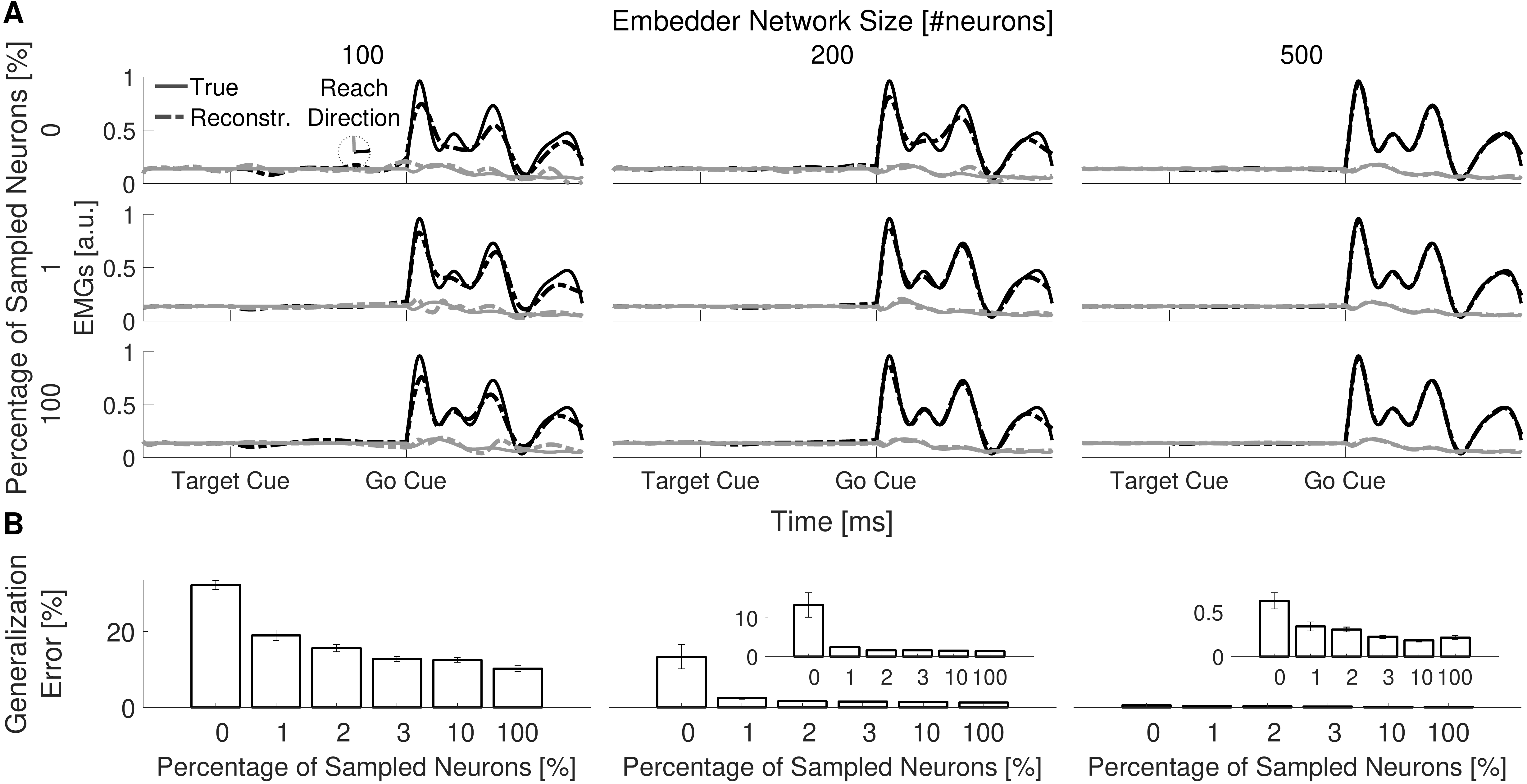}
\squeezeupCap
\caption{\textbf{Ground-truth and reconstructed EMGs} \textbf{(A)} Dashed lines: reconstructed EMGs of muscle 1 generated by embedder networks of different sizes, and trained using different percentages of ground-truth neurons; solid lines: respective ground-truth EMGs (generated by the biological network). \textbf{(B)} Generalization error averaged across muscles and reach directions. Insets: magnifications of the respective plots.} 
\label{fig3}
\end{figure} 

\begin{figure}[t]
\includegraphics[width=\textwidth]{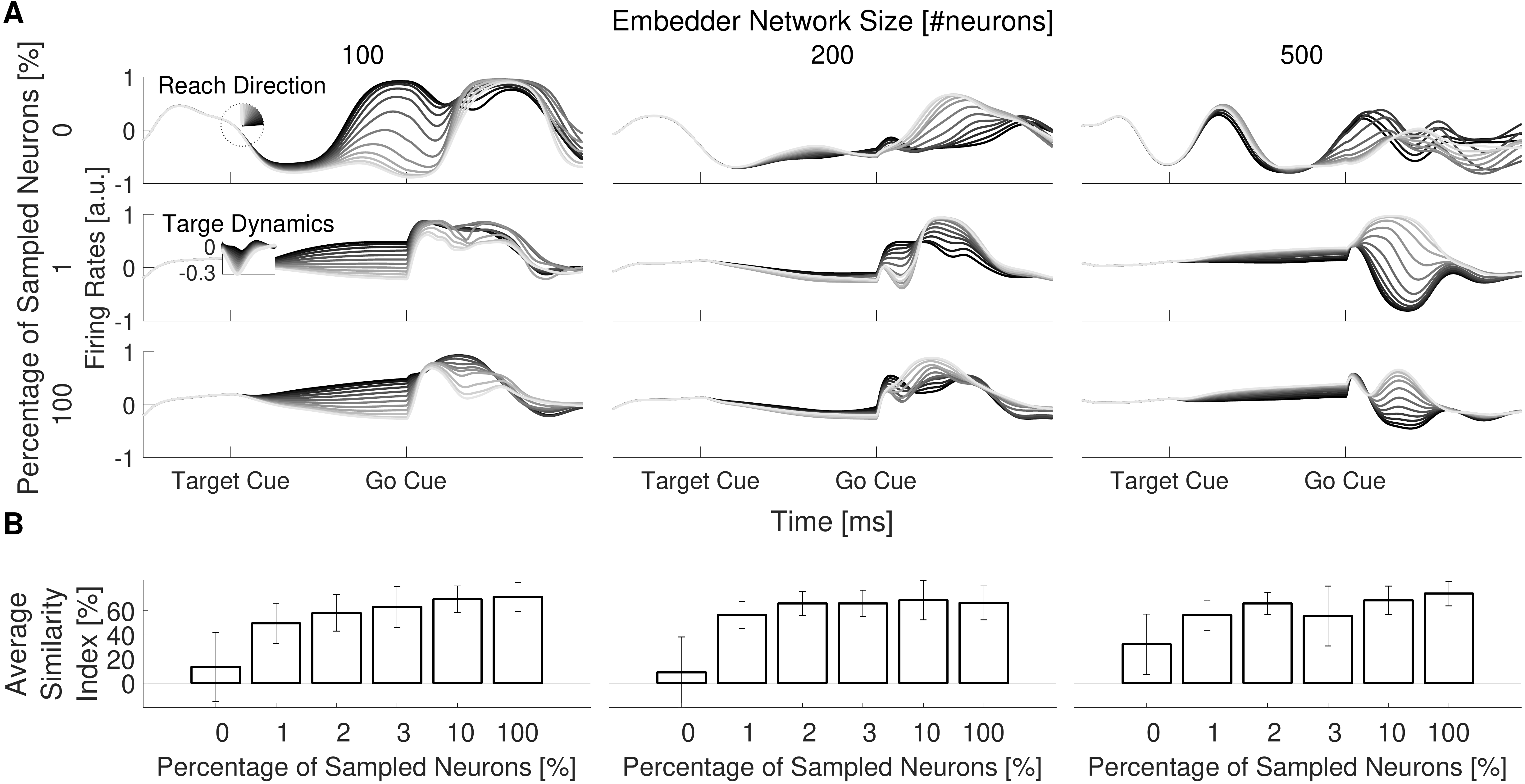}
\squeezeupCap
\caption{\textbf{Neural activities and Activation Similarity Index of embedder networks}. \textbf{(A)} Firing rates of one representative neuron belonging to embedder networks of different sizes, and trained using different percentages of ground-truth neurons. Traces are color coded based on target direction. Inset: target firing rates of an example neuron specified by the biological network. \textbf{(B)} Average Similarity Index as a function of the percentage of neurons sampled from the biological network.}
\label{fig4}
\end{figure}

\section{Results}
\subsubsection{Embedding the Dynamics Improves RNN ability to Generalize.} To assess how well the emebedder networks learned to perform the task, we measured their ability to approximate the EMG signals generated by the biological network for the test reach directions. Figure \ref{fig3}A shows representative EMGs generated by the biological and embedder networks, for different combinations of network size ($\eta$) and percentage of sampled neurons ($\gamma$). Figure \ref{fig3}B shows the average generalization error (GE) for all the trained embedder networks. 
\\ \indent All the trained RNNs reproduced the target EMGs with relatively high accuracy ($GE\leq32.25\%$), and their performance improved with embedder network size ($\eta$). Critically, the generalization error of the RNNs that were only trained to generate the appropriate input-output behavior was significantly higher than that of the
RNNs that were also trained to internally generate the activation dynamics of the biological network ($p=1.26$ x $10^{-35}$, Fig. \ref{fig5}A). Finally, GE decreases with $\gamma$ ($p=6.25$ x $10^{-4}$), although only by a small amount ($\beta_1 = -1.94$ x $10^-2$, Fig. \ref{fig5}B). Therefore, embedding the activation dynamics of the biological network improves the ability of the embedder networks to generalize.
\squeezeup
\subsubsection{Embedding the Dynamics Increases RNN Activation Similarity.} 
\begin{figure}[t]
\includegraphics[width=\textwidth]{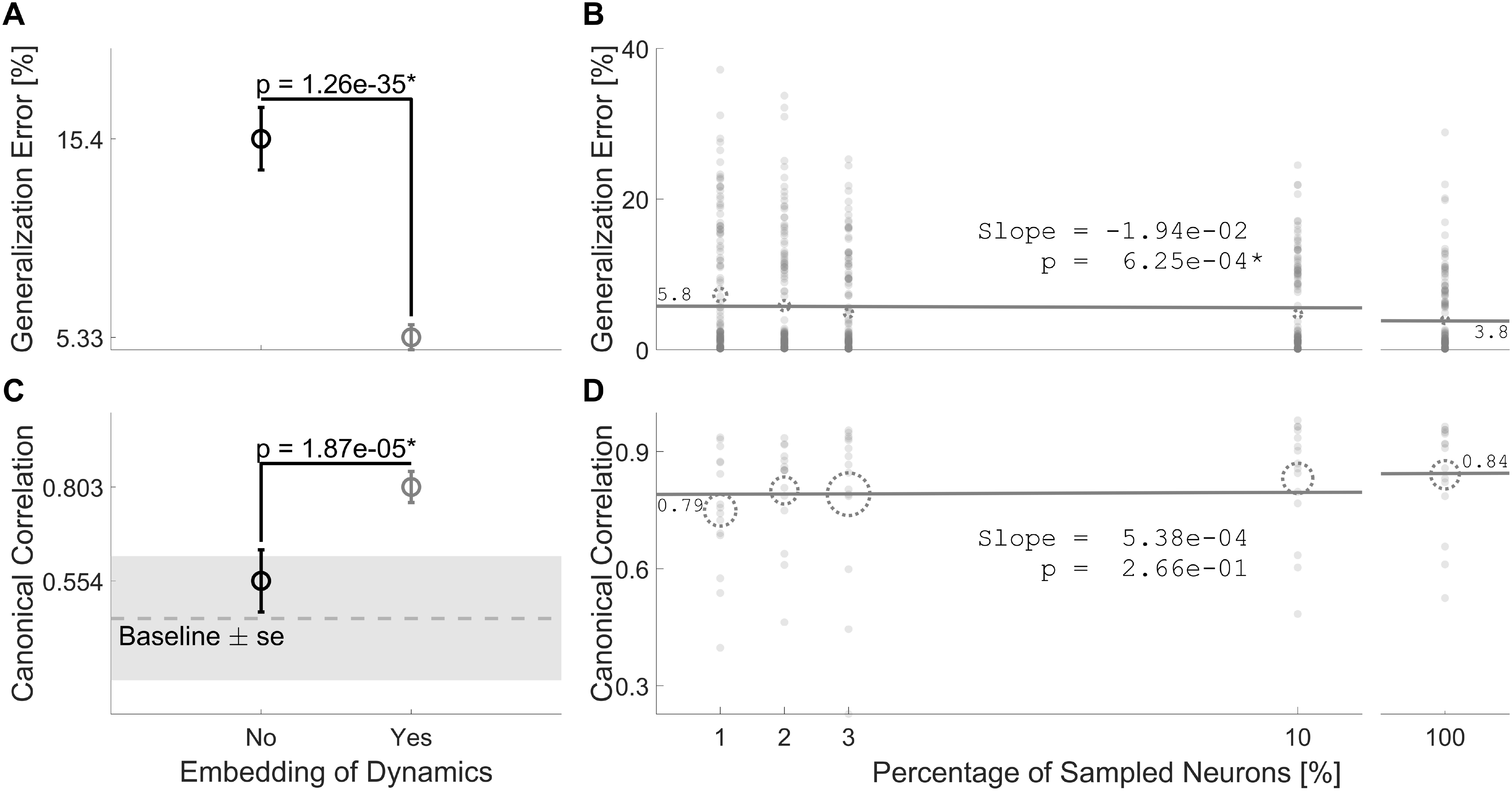}
\squeezeupCap
\caption{\textbf{Generalization error and canonical correlations} \textbf{(A)} Generalization error (GE) averaged across muscles, reach directions, and networks, for networks trained with (grey) and without (black) embedding of dynamics. \textbf{(B)} GE as a function of the percentage of sampled neurons ($\gamma$). Filled circles: individual muscles; dotted circles: average across muscles and networks; line: linear fit. \textbf{(C)} Canonical correlations (CC) averaged across canonical variables and networks, for networks trained with (grey) and without (black) embedding of dynamics. Baseline: average CC between baseline and biological network. \textbf{(D)} CC as a function of $\gamma$. Filled circles: individual canonical correlation; dotted circles: average across CC and networks; line: linear fit. Radiuses of dotted circles in panels (B) and (D) represent standard errors.}
\label{fig5}
\end{figure}

To quantify the similarity between the activation dynamics of biological and embedder networks, we computed their singular vector canonical correlations \cite{raghu2017svcca}. Figure \ref{fig4}A shows the firing rates of a representative unit across selected reach directions, for different combinations of $\eta$ and $\gamma$. Note that when the RNN dynamics were {\it constrained} to be similar to those of the biological network ($\gamma>0$), RNN units show activation patterns that resemble those observed in the motor cortex neurons during reaching: slow ramping activity during the preparatory phase followed by fast oscillations during the movement phase. Figure \ref{fig4}B shows the average activation similarity index (ASI) for all the trained embedder networks. Clearly, the ASI of the RNNs trained with $\gamma=0$ approaches $0\%$; this suggests that untrained RNNs and RNNs trained to only reproduce the input-output behaviour of the biological network have activation dynamics that are comparatively dissimilar to those of the biological network. 
\\ \indent Critically, the canonical correlations (CC) of these input-output trained RNNs were significantly lower than those of the RNNs that were also trained with embedding of dynamics ($p=1.87$ x $10^{-5}$, Fig. \ref{fig5}C). Importantly, CC do not significantly increase with $\gamma$ ($p=2.66$ x $10^{-1}$, Fig. \ref{fig5}D). This suggests that constraining the dynamics using the activation patterns of only $1\%$ of neurons of the biological network is enough to have the RNN capture the dominant modes of shared variability. Therefore, embedding the activation dynamics using a very small activity sample is enough to have the embedder network capture important response properties of the entire biological network. The remarkable amount of information contained in the response properties of such a small activity sample is consistent with what was observed in previous experimental \cite{williamson2016scaling} and theoretical \cite{gao2017theory} work.

\section{Conclusion}

A common strategy for the modeling of biological neural systems exploiting RNNs is the so-called {\it task-based} approach \cite{sussillo2014neural}. This method trains the RNNs to learn the mapping between the input and the output signals of the biological circuit of interest.
However, the problem of identifying RNN parameters from input-output data is generally ill-posed, so that there can be manifolds of equivalent solutions in the parameters space. As a consequence, even though RNNs generally manage to approximate the input-output behavior accurately, the learned internal computations and activity dynamics might differ substantially
from the ones of the target biological system \cite{williamson2019bridging}.
\\ \indent
In this work, we proposed a novel training strategy, which allows learning the input-output behavior of an RNN together with its internal network dynamics, based on sparse neural recordings.
We  tested  the  proposed  method  by  training an RNN to simultaneously reproduce the internal dynamics and output signals of a physiologically-inspired neural model for the generation of the motor cortical and muscle activation dynamics typically observed during reaching movements.
\\ \indent
Critically, we demonstrated our approach to be successful even in the presence of small samples of activation dynamics from the target neural model. 
Furthermore, we showed that the RNNs trained with this approach achieved superior generalization performance. In sum, these results suggest that our method is suitable for the automatic generation of RNN models that capture critical computational properties of the neural circuit of interests. The resulting RNN models can then be further analyzed with methods from dynamical systems theory \cite{sussillo2013opening} in order to reverse engineer the computational principles of brain function. 

%
%
%
\bibliographystyle{splncs04}
\bibliography{bibliography_AS_ICANN20}

\end{document}